\documentclass[conference]{IEEEtran}
\IEEEoverridecommandlockouts

\usepackage{amsmath,graphicx}

\usepackage{amsfonts} 
\usepackage{empheq} 
\usepackage{color}
\usepackage{mathrsfs}
\usepackage{multirow,cite}
\usepackage{comment} 
\usepackage{bm} 

\usepackage{bbding}
\usepackage{pifont}
\usepackage{wasysym}
\usepackage{amssymb}

\allowdisplaybreaks[4]

\usepackage{booktabs}

\newcommand{\ie}{{i.e.,~}}
\newcommand{\eg}{{e.g.,~}}

\newcommand{\aka}{{a.k.a.~}}
\newcommand{\etal}{{et~al.}}

\usepackage{algorithmic}
\usepackage{graphicx}
\usepackage{textcomp}
\usepackage{xcolor}
\def\BibTeX{{\rm B\kern-.05em{\sc i\kern-.025em b}\kern-.08em
    T\kern-.1667em\lower.7ex\hbox{E}\kern-.125emX}}
\begin{document}
\newcommand{\red}[1]{\textcolor{red}{#1}}

\title{Audio-Visual Speech Enhancement \\ With Selective  Off-Screen Speech Extraction\\
\thanks{*These two authors contributed equally to this work.
This work is supported in part by  JSPS KAKENHI Nos. 19H04137, 21H05054, 22J22424, 22KJ2959.
We thank Mr. Masaki Kuribayashi for his valuable feedback.}
}


\author{Tomoya Yoshinaga$^{1*}$, Keitaro Tanaka$^{1*}$, Shigeo Morishima$^{2}$  \\
$^{1}$ School of Advanced Science and Engineering, Waseda University, Tokyo, Japan \\
$^{2}$ Waseda Research Institute for Science and Engineering, Tokyo, Japan}

\maketitle

\begin{abstract}
\noindent
This paper describes an audio-visual speech enhancement (AV-SE) method that estimates from noisy input audio a mixture of the speech of the speaker appearing in an input video (on-screen target speech) and of a selected speaker not appearing in the video (off-screen target speech).
Although conventional AV-SE methods have suppressed all off-screen sounds, it is necessary to listen to a specific pre-known speaker's speech (\eg family member's voice and announcements in stations) in future applications of AV-SE (\eg hearing aids), even when users' sight does not capture the speaker.
To overcome this limitation, we extract a visual clue for the on-screen target speech from the input video and a voiceprint clue for the off-screen one from a pre-recorded speech of the speaker.
Two clues from different domains are integrated as an audio-visual clue,
and the proposed model directly estimates the target mixture.
To improve the estimation accuracy, we introduce a temporal attention mechanism for the voiceprint clue and propose a training strategy called the muting strategy.
Experimental results show that our method outperforms a baseline method that uses the state-of-the-art AV-SE and speaker extraction methods individually in terms of estimation accuracy and computational efficiency.
\end{abstract}


\section{Introduction}
\label{sec:introduction}
\vspace{-0.2mm}

Audio-visual speech enhancement (AV-SE) aims to extract a target speaker's speech from a noisy input signal (\aka speech enhancement) by using an additional visual clue of the speaker, typically lip movement~\cite{conversation,cochleanet}.
As lip movement 
 helps us to track the synchronizing target speech 
 contaminated by non-speech noise or interfering speech,
 AV-SE works robustly to various kinds of noise.
AV-SE has the potential for practical applications,
 such as hearing aids~\cite{cochleanet, overview}, telecommunication~\cite{codecs},
 and automatic speech recognition front end~\cite{look-to-listen}.


The standard approach of AV-SE is to extract only the speech of the speaker appearing in an input video (on-screen target speech) and suppress the other sounds (off-screen sounds).
Based on the fact that humans improve their speech perception by watching a speaker's face~\cite{perception}, 
a pioneer study on AV-SE~\cite{pioneer} applied a statistical speech enhancement model that used lip shape features.
Since deep neural networks (DNNs) appeared,
 DNN-based methods~\cite{conversation,look-to-listen} have been the major approach of AV-SE because of DNNs' high capability of extracting speech and fusing multimodal information.
Recently, several studies have also tackled practical problem settings such as temporal occlusion of the speaker's mouth~\cite{conceal, AVSE+speaker_extraction}.

In some practical situations, however, users need to listen to a pre-known speaker's speech in the off-screen sounds as well as the on-screen target speech.
For example, when AV-SE is applied to hearing aids that extract an interlocutor's speech, 
young users should always pay attention to 
what their parents or teachers say,
even when users' sight does not capture the speaker's face.
Announcements in stations also should not be suppressed for user safety.
These situations call for a method 
 that can simultaneously extract the on-screen target speech 
 and selected off-screen speech (off-screen target speech)
 from a noisy input signal, 
 where the voice characteristics of the off-screen target speaker are pre-known.

\begin{figure}
  \centering
  \centerline{\includegraphics[width=0.99\columnwidth]{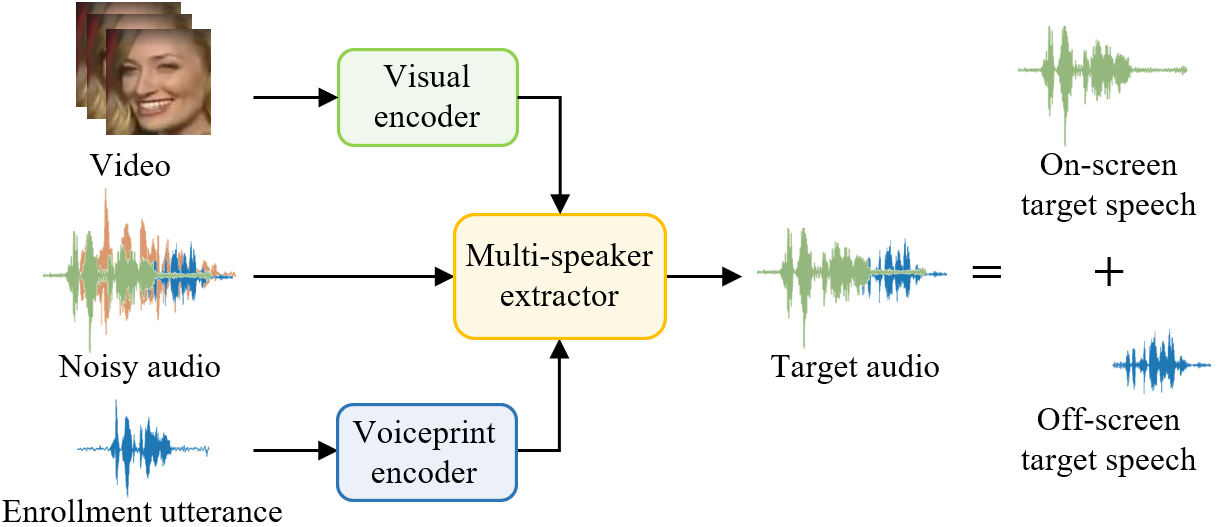}}
\vspace{-1mm}
\caption{
    The proposed method directly extracts the mixture of on-screen and off-screen target speech from noisy audio using the corresponding video and enrollment utterance.
}
\label{fig:teaser}
\vspace{-3mm}
\end{figure}
A straightforward approach to this situation is
 to extract each target speech individually and mix them.
The on-screen target speech can be extracted by AV-SE with the help of the corresponding visual clue.
The off-screen one can be extracted by speaker extraction~\cite{voicefilter, spex++} with the help of a voiceprint clue from a pre-recorded speech of the speaker (enrollment utterance).
However, the mixing operation 
deteriorates the output signals
 because of the accumulation of estimation errors, 
 such as artifacts and remaining non-speech noises or interfering speech in each output.
This approach is also computationally inefficient because two independent models
 are required to obtain each output
 from the same input audio,
 which would be undesirable in low-resource devices~\cite{multi-target}. 


In this paper,
 we propose a unified AV-SE model with selective off-screen speech extraction
 that directly estimates a mixture of the on-screen and off-screen target speech (Fig.~\ref{fig:teaser}).
At the heart of our method is the multimodal fusion of two clues obtained from different domains.
Specifically, we extract the visual clue for the on-screen target speech from the input video
 and the voiceprint clue for the off-screen target speech from the enrollment utterance.
These clues are integrated as an audio-visual clue,
 with which a multi-speaker extractor extracts the target mixture from the input audio.
End-to-end training enables our model to perform more accurately than the mixing-based approach.

The main contribution of this study is to propose a computationally efficient yet high-performance method for a novel practical problem setting in AV-SE.
Our model further improves its performance by attention mechanism and muting strategy.
When the off-screen target speech temporally does not exist in the noisy input audio,
 the model does not need to refer to the voiceprint clue.
We thus estimate the voice activity of the off-screen target speech
 and calculate temporal attention,
 which controls how much the voiceprint clue contributes to the audio-visual clue.
We also utilize a muting strategy, 
 where either on-screen or off-screen target speech in the noisy input audio is muted during training.
This encourages the model 
 to strongly bind each clue and the characteristics of the corresponding signal.
Experimental results show that 
 our method outperforms a straightforward baseline method (even the combination of the state-of-the-art AV-SE and speaker extraction methods)
 in terms of the quality of output signals and the number of model parameters.


\section{Related work}
\label{sec:related work}


This section reviews target speaker extraction methods in terms of the modalities of their clues. We also briefly consider denoising methods as alternative approaches to our problem.

\vspace{-0.5mm}
\subsection{Target speaker extraction}
\label{ssec:target_speaker_extraction}
\vspace{-0.5mm}

Target speaker extraction aims to extract the target speech 
 from a noisy input signal using additional information about the speaker. 
Existing methods can mainly be categorized into two approaches: 
 visual-clue-based and audio-clue-based. 
The visual-clue-based approach, namely AV-SE, 
 utilizes lip movements~\cite{conversation,cochleanet} 
 or crops of face images~\cite{look-to-listen} synchronized with the target speech. 
On the other hand, the audio-clue-based approach, namely speaker extraction, 
 utilizes a speaker-dependent voiceprint~\cite{voicefilter, spex++} obtained from an enrollment utterance. 
Although these approaches have flourished independently, 
 some studies are recently trying to integrate them as an audio-visual-clue-based approach. 
In this approach, the visual and voiceprint clues can work complementarily, 
 and thus models become robust 
 against temporal occlusion of lip movements~\cite{conceal, AVSE+speaker_extraction} 
 or contamination in enrollment utterances~\cite{AVSE+speaker_extraction}.
Note that both clues are designed to extract the same on-screen target speech.
Our method is also one of the audio-visual-clue-based approaches,
 but we use the two clues to extract different target speech signals.

\vspace{-0.5mm}
\subsection{Denoising}
\label{ssec:selective_noise_suppression}
\vspace{-0.5mm}

While the proposed method estimates the target mixture
 by additively extracting two speech signals from the input audio,
 our goal might be achieved by subtractively suppressing sounds
 other than the two speech signals contained in the input audio.
Such approaches are called selective noise suppression~\cite{n-hans}
 or audio-only speech enhancement~\cite{segan, dccrn}.
Selective noise suppression removes only unnecessary noises
using their enrollment recordings 
 without removing \textit{necessary noises} (\eg alarms), and
 audio-only speech enhancement removes all non-speech sounds uniformly.
However, these approaches are actually not appropriate for our goal
 because they require all enrollment recordings of unnecessary noises
 prepared in advance 
 or fail to remove unnecessary speech.
In contrast, our additive approach requires only one enrollment utterance in advance
 to work in any unknown acoustic conditions.

\section{Proposed method}
\label{sec:proposed_method}

This section explains the proposed framework that directly extracts the mixture of the on-screen and off-screen target speech from the noisy input audio.
We also present the attention mechanism and muting strategy that are uniquely motivated by the framework to improve the model's performance in multi-target and multi-modal speaker extraction.

\subsection{Framework}
\vspace{-1mm}
\label{ssec:framework}

\begin{figure*}[!t]
 \begin{tabular}{cc}
  \begin{minipage}[b]{0.47\linewidth}
   \centering
   \includegraphics[keepaspectratio, width=0.99\columnwidth]{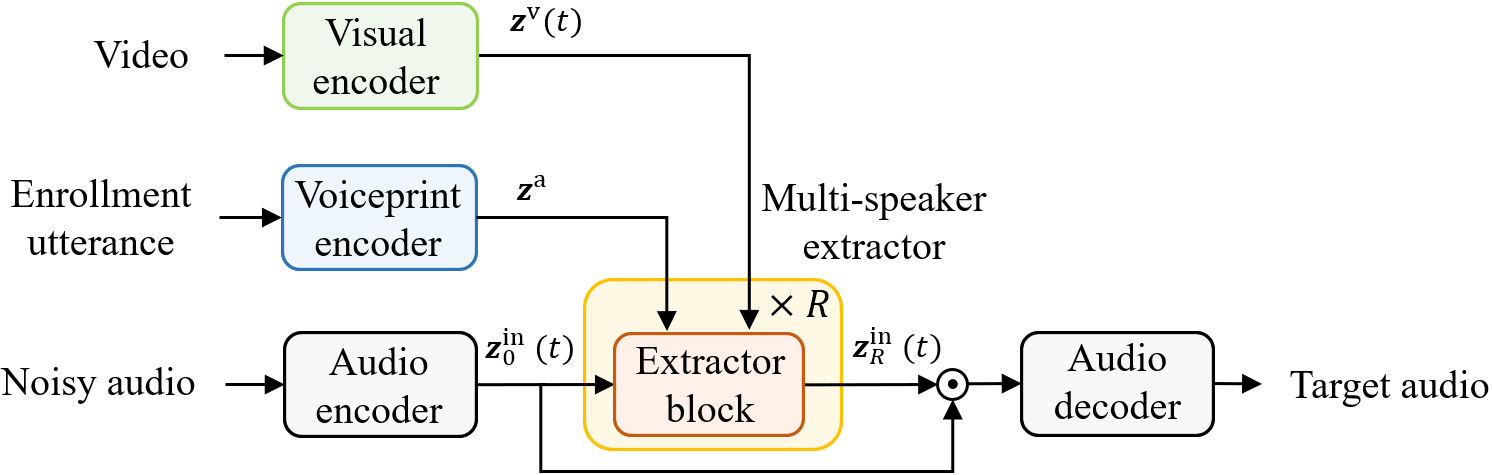}
   \centerline{\fontsize{8pt}{0cm}\selectfont (a) Overall architecture of our model.}\medskip
  \end{minipage} &

  \begin{minipage}[b]{0.47\linewidth}
   \centering
   \includegraphics[keepaspectratio, width=0.99\columnwidth]{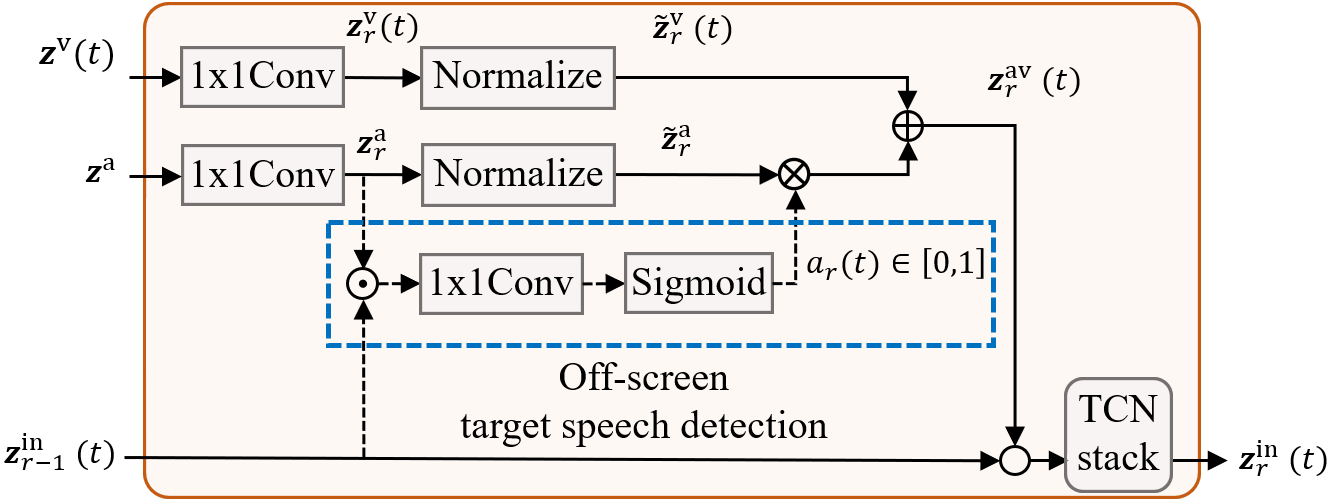}
   \centerline{\fontsize{8pt}{0cm}\selectfont (b) The $r$-th extractor block.}\medskip
  \end{minipage}
 \end{tabular}
\vspace{-3.5mm}
\caption{
Illustration of our model. 
$\odot$, $\otimes$, $\oplus$, and $\ocircle$ represent element-wise multiplication, multiplication between a vector and scalar, summation, and concatenation, respectively.
(a) Overall architecture based on Conv-TasNet~\cite{convtasnet}. 
(b) Details of the $r$-th extractor block. The dotted arrows are only performed when we use the attention mechanism.
}
\label{fig:concept}
\vspace{-3mm}
\end{figure*}

 Our model is a time-domain encoder-decoder model overall (Fig.~\ref{fig:concept}(a)).
It consists of five parts: audio encoder, visual encoder, voiceprint encoder, multi-speaker extractor, and audio decoder.
The audio encoder transforms the noisy input audio waveform 
 into initial time-variant latent representations $\mathbf{z}^{\mathrm{in}}_{0}(t) \in \mathbb{R}^{D^{\mathrm{in}}} (t = 1, ..., T)$,
 where $D^{\mathrm{in}}$ and $T$ are the dimension of each latent representation and the number of time frames, respectively.
The visual encoder extracts time-variant embeddings 
$\mathbf{z}^{\mathrm{v}}(t) \in \mathbb{R}^{D^\mathrm{v}}$ from a sequence of cropped lip images in the input video synchronized with the on-screen target speech,
 where $D^{\mathrm{v}}$ is the dimension of each embedding.
Note that the last layer of the encoder conducts upsampling to match the temporal resolution of $\mathbf{z}^{\mathrm{in}}_{0}(t)$.
The voiceprint encoder extracts a time-invariant speaker embedding $\mathbf{z}^{\mathrm{a}} \in \mathbb{R}^{D^\mathrm{a}}$ from the enrollment utterance to represent the voice characteristics of the off-screen target speaker (\eg pitch and timbre),
 where $D^{\mathrm{a}}$ is the dimension of the embedding.
Given $\mathbf{z}^{\mathrm{v}}(t)$, $\mathbf{z}^{\mathrm{a}}$, and $\mathbf{z}^{\mathrm{in}}_{0}(t)$,
 the multi-speaker extractor calculates time-variant masks $\mathbf{z}^{\mathrm{in}}_{R}(t) \in \mathbb{R}^{D^{\mathrm{in}}}$ 
 which are applied to $\mathbf{z}^{\mathrm{in}}_{0}(t)$,
 where $R$ is the number of iterations (explained below).
Then, the audio decoder estimates the target audio from the masked initial latent representations of the input mixture.

For the visual and voiceprint encoder, we use the corresponding encoding modules of existing AV-SE and speaker extraction models, respectively.
The other three parts are based on a time-domain speech separation model Conv-TasNet~\cite{convtasnet}.
Below, we focus on the multi-speaker extractor, 
which is the heart of the proposed framework.
The multi-speaker extractor consists of $R$ consecutive extractor blocks (Fig.~\ref{fig:concept}(b)) and they
form $\mathbf{z}^{\mathrm{in}}_{R}(t)$
 by iteratively processing $\mathbf{z}^{\mathrm{in}}_{0}(t)$ conditioned by an audio-visual clue instead of a single-modal clue as in AV-SE or speaker extraction.
Let $r$ be an iteration index $(r = 1 ,..., R)$.
$\mathbf{z}^{\mathrm{v}}(t)$ and $\mathbf{z}^{\mathrm{a}}$ are individually processed by pointwise convolution layers
 and transformed to ${\mathbf{z}}^{\mathrm{v}}_r(t) \! \in \! \mathbb{R}^{D^{\mathrm{av}}}$ 
 and ${\mathbf{z}}^{\mathrm{a}}_r \! \in \! \mathbb{R}^{D^{\mathrm{av}}}$,
 where $D^{\mathrm{av}}$ is the dimension of each embedding.
We sum their normalized embeddings, 
$\tilde{\mathbf{z}}^{\mathrm{v}}_r(t)$ and $\tilde{\mathbf{z}}^{\mathrm{a}}_r$,
 and obtain an audio-visual embedding ${\mathbf{z}}^{\mathrm{av}}_r(t) \in \mathbb{R}^{D^{\mathrm{av}}}$.
${\mathbf{z}}^{\mathrm{av}}_r(t)$ is concatenated with $\mathbf{z}^{\mathrm{in}}_{r-1}(t)$
 and processed by temporal convolutional network (TCN) stack~\cite{reentry}.
The output of the TCN stack, $\mathbf{z}^{\mathrm{in}}_{r}(t)$, is used for the next iteration.
After $R$ iterations, 
 the initial latent representation $\mathbf{z}^{\mathrm{in}}_{0}(t)$ is multiplied 
 by the estimated mask $\mathbf{z}^{\mathrm{in}}_{R}(t)$ element-wisely
 and put into the audio decoder.
The entire model is trained end-to-end so that a scale-dependent signal-to-noise ratio (SNR)~\cite{sdr} is maximized:
\begin{align}
    {\mathrm{SNR}}_{\mathrm{dB}}
    = -\mathcal{L}_{\mathrm{on}+\mathrm{off}} 
    = 10 \log_{10}\frac{\| s \| ^2}{\| \hat{s} - s \|^2},
\end{align}
 where $s$ and $\hat{s}$ are the clean and estimated signals, respectively.

When we consider extracting each target speech individually and mixing them,
 one of the main drawbacks is the accumulation of estimation errors, such as artifacts and remaining non-speech noises or interfering speech.
In contrast, since the proposed model extracts both target speech signals in a lump, 
 we can avoid this problem and further conduct end-to-end training.
The proposed model is also computationally efficient
 because, unlike the mixing-based approach,  we need only one model to obtain the output mixture.

\vspace{-2mm}
\subsection{Attention mechanism}
\vspace{-1mm}
\label{ssec:attention}

While the visual clue is time-variant and depends on (\ie synchronized with) the on-screen target speech~\cite{conversation},
 the voiceprint clue is time-invariant and independent of (\ie not synchronized with) the off-screen target speech~\cite{voicefilter}.
To fill this gap, we prompt the multi-speaker extractor to refer to the voiceprint clue 
 only when the off-screen target speech temporally exists in the noisy input audio
 by the attention mechanism~\cite{facefilter}.
 Specifically, we use the speaker-dependent voice activity detection (SDVAD) network~\cite{wase},
 which takes noisy speech as an input and estimates whether the target speaker is active at each frame using the speaker embedding.

The off-screen target speech detection network for the SDVAD method takes as an input the element-wise multiplication 
 between $\mathbf{z}^{\mathrm{a}}_r$ and $\mathbf{z}^{\mathrm{in}}_{r-1}(t)$
 and estimates $a_r(t)\in[0,1]$ at each time frame as shown in Fig.~\ref{fig:concept}(b).
$a_r(t)$ represents the confidence that 
 $t$-th frame of the noisy input contains the active off-screen target speech.
We utilize $a_r(t)$ as an attention
 that controls the contribution of $\tilde{\mathbf{z}}^{\mathrm{a}}_r$ to ${\mathbf{z}}^{\mathrm{av}}_r(t)$ as follows:
\begin{align}
    {\mathbf{z}}^{\mathrm{av}}_r(t) = \tilde{\mathbf{z}}^{\mathrm{v}}_r(t) + a_r(t) \tilde{\mathbf{z}}^{\mathrm{a}}_r.
\end{align}
Following~\cite{wase}, 
 the off-screen target speech detection network is trained such that 
 the cross-entropy $\mathcal{L}_{\mathrm{CE}}$ between $a_r(t)$ and the oracle voice activity 
 ($1$ if the speech exists and $0$ otherwise) is minimized.
The entire model is trained in a multi-task learning manner under 
 the total loss function $\mathcal{L}_{\mathrm{total}}$:
\begin{align}
    \mathcal{L}_{\mathrm{total}} = \mathcal{L}_{\mathrm{on}+\mathrm{off}}  + \lambda \mathcal{L}_{\mathrm{CE}},
\end{align}
 where $\lambda$ is a hyperparameter to control the weight of the cross-entropy loss function.

\subsection{Muting strategy}
\label{ssec:training}

To further improve the performance of our model,
 we introduce a novel training strategy, which we call the muting strategy.
While the proposed model refers to two clues from different domains,
 it outputs a mixture of two speech signals without distinction.
 Thus the model does not explicitly interpret the correspondence
 between each clue and the characteristics of the target speech of the clue.
Each clue can further contribute to extracting the precise target speech by encouraging the model to learn the correspondence more precisely.
 We, therefore, mute one of the on-screen or off-screen target speech signals at certain probabilities during training
 (at $p_{\mathrm{on}}$ for on-screen and $p_{\mathrm{off}}$ for off-screen).
Since this forces the model to output only either of the two target speech signals, 
the model strongly binds each clue and the characteristics of the corresponding signal.


\section{Evaluation}
\label{sec:evaluation}

This section describes experiments to evaluate the performance of our method for multi-speaker extraction. We consider four conditions where the target mixture is contaminated by environmental sound noise and/or interfering speech. 

\vspace{-2mm}
\subsection{Data}
\vspace{-1mm}
\label{ssec:data}

{\tabcolsep=3mm
\begin{table*}[t]
\centering
\caption{Comparison of the baseline method and the proposed method on the conditions with environmental sound noise (``noise") and with an interfering speaker (``spk"). 
AM and MS refer to the attention mechanism and muting strategy, respectively.}
\label{tab:ablation}
\vspace{-2.5mm}
\begin{tabular}{cccp{3mm}ccccp{3mm}c}
\toprule
\multirow{3}{*}{Methods} 
 & \multirow{3}{*}{AM} & \multirow{3}{*}{MS} &
 & \multicolumn{2}{c}{``noise" condition} & 
 \multicolumn{2}{c}{``spk" condition} & &
 \multirow{3}{*}{\#Params $\downarrow$}\\
 \cmidrule(lr){5-6} \cmidrule(lr){7-8}
 &&&& SI-SDRi (dB) $\uparrow$ & SDRi (dB) $\uparrow$&
 SI-SDRi (dB) $\uparrow$ & SDRi (dB) $\uparrow$ &&\\
\midrule
Baseline\_A~\cite{muse, wase} &  - & -  &
& 7.34 & 7.19  & 7.58 & 7.96 && 29.8M 
\\
\midrule

\multirow{4}{*}{Proposed\_A} & - & - && 7.56 & 7.37 & 8.25 & 8.56 && 25.1M\\
& - & \checkmark && 7.67 & 7.47 & 8.44 & 8.78 && 25.1M\\
& \checkmark & - && 7.77 & 7.57 & 8.29 & 8.56 && 25.1M\\
& \checkmark  & \checkmark &
& 8.06 & 7.88 & 8.73 & 9.11 && 25.1M \\
\midrule
Baseline\_r~\cite{reentry, spex++} &  - & -  &
& 7.78 & 7.65 & 8.22 & 8.60  && 53.0M 
\\
\midrule
\multirow{4}{*}{Proposed\_r} & - & - && 8.40 & 8.19 & 9.49 & 9.77  && \textbf{19.2M}\\
& - & \checkmark &&  8.55 & 8.37 & 9.80 & 10.10 && \textbf{19.2M}\\
& \checkmark & - && 8.42 & 8.21 & 9.71 & 9.98 && \textbf{19.2M}\\
& \checkmark  & \checkmark && \textbf{8.58} & \textbf{8.41} & \textbf{9.91} & \textbf{10.21} && \textbf{19.2M}\\
\bottomrule
\end{tabular}
\vspace{1mm}
\\
\hspace{110mm}
$\uparrow$ means higher is better, and $\downarrow$ means lower is better.
\vspace{-5mm}
\end{table*}
}

We used the VoxCeleb2 dataset~\cite{voxceleb2}, WSJ0 corpus~\cite{wsj0}, and AudioSet~\cite{audioset} for the on-screen target speech, off-screen target speech, and environmental sound noise, respectively.
The VoxCeleb2 consists of speech signals and their synchronizing videos of the speaker's face region,
 while WSJ0 consists of only speech signals labeled with speaker identity.
AudioSet contains audio clips labeled with multiple classes (527 classes in total), 
 such as human voices, music, and sounds of things. 
For the training and validation, we used 25,000 (800 speakers), 12,776 (101 speakers), and 18,870 clips 
 in the VoxCeleb2 training set, WSJ0 ``si\_tr\_s” set, and the balanced training subset of the AudioSet, respectively.
Each set was split into a training set ($80\%$) and a validation set ($20\%$).
For the test, we used 3,000 (118 speakers), 1,857 (18 speakers), and 3,000 clips 
 in the VoxCeleb2 test set, WSJ0 ``si\_dt\_05” and ``si\_et\_05” sets, and the evaluation subset of the AudioSet, respectively.
Note that, in the VoxCeleb2 and WSJ0, speakers in the test set were unseen in the training and validation set. 

With these datasets, 
 we generated 20,000, 5,000, and 3,000 pairs of noisy input audio and the synchronizing video
 for the training, validation, and test sets.
All sounds and videos were resampled to 16 kHz and 25 fps.
We generated the input audio 
 by mixing an off-screen target speech and a four-second environmental sound noise 
 into a four-second on-screen target speech 
 with random SNR between -2.5 and 2.5 dB.
The off-screen target speech was cropped to have a random duration between two and four seconds for the training and validation 
 and between zero and four seconds for the test.
We randomly selected the enrollment utterance from the other utterances of the off-screen target speaker.
We also conducted an experiment where the target mixture is contaminated by another speaker (``spk") instead of AudioSet noise (``noise").
Moreover, experiments with both AudioSet noise and an interfering speaker (``noise+spk") and with two interfering speakers (``2spk") were conducted.
In these situations, interfering speech was randomly selected from the VoxCeleb2 or WSJ0.

\vspace{-0.5mm}
\subsection{Model configuration}
\vspace{-0.7mm}
\label{ssec:configuration}

For the visual encoder, we used the visual encoder of AV-ConvTasnet~\cite{muse} (Proposed\_A)
 or the attractor encoder of the reentry~\cite{reentry} (Proposed\_r), 
with $D^{\mathrm{v}}$ of 512 or 256, respectively.
Note that the attractor encoder takes the additional noisy audio as an input as well as the video.
In our implementation, AV-ConvTasnet was a slightly modified version~\cite{muse} of the original~\cite{av-convtasnet}.
For the voiceprint encoder, we used WASE~\cite{wase}, where 
$D^{\mathrm{a}}$ was 256.
Finally, we used AV-ConvTasnet~\cite{muse} for the audio encoder, TCN stack, and audio decoder.
The parameters of the extractor block $D^{\mathrm{av}}$, $D^{\mathrm{in}}$, and $R$ were set to 256, 256, and 4, respectively.
$\lambda$, $p_{\mathrm{on}}$, and $p_{\mathrm{off}}$ were set to 1.0, 20\%, and 20\%, respectively.

We trained all models with an Adam 
optimizer for 200 epochs.
The learning rate was initialized to $0.001$
 and halved if the validation loss did not improve for three epochs.
Early stopping was applied if the learning rate dropped four times.
Note that the optimizations for Proposed\_A and Proposed\_r follow the methods used in AV-ConvTasnet and the reentry, respectively.
Specifically, for Proposed\_A, we optimized our entire model except for the pre-trained visual encoder's front-end~\cite{visual-frontend}, which consists of a 3D convolutional layer and ResNet18 block.
For Proposed\_r, we optimized our entire model except for the SLSyn network~\cite{reentry} of the visual encoder, which was pre-trained to extract speech-lip synchronization embeddings in reentry, and then fine-tuned the entire model with the re-initialized optimizer.

\vspace{-0.3mm}
\subsection{Evaluation metrics}
\vspace{-0.5mm}
\label{ssec:metrics}

We evaluated our method in terms of 
 estimation accuracy and computational efficiency.
We did the estimation accuracy 
 for the multi-speaker extraction using scale-invariant signal-to-distortion ratio~\cite{si-sdr} improvement (SI-SDRi) 
 and signal-to-distortion ratio~\cite{sdr} improvement (SDRi), measuring the amount of distortion in estimated signals.
We did the computational efficiency 
 by the number of model parameters.

\vspace{-0.3mm}
\subsection{Baseline methods}
\vspace{-0.5mm}
\label{ssec:metrics}

We set two baseline methods,
 which individually extract the on-screen and off-screen target speech
 using AV-SE and speaker extraction, respectively, and then mix them.
 Specifically, we used the combination of AV-ConvTasnet and WASE (Baseline\_A) and that of the reentry and SpEx++~\cite{spex++} (Baseline\_r). 
 We selected AV-ConvTasnet and WASE because their network and Propoed\_A consist of Conv-TasNet conditioned with a clue (note that only WASE has the skip-connection paths~\cite{convtasnet}).
This enables a fair comparison between the direct and separative approaches at a framework-level since the audio-visual clue is obtained with the same encoders as those of the direct one. 
The reentry and SpEx++ are state-of-the-art in each task.
These AV-SE and speaker extraction methods were originally evaluated on the VoxCeleb2 and WSJ0, respectively.
Thus we used the datasets not to deteriorate the two baseline methods rather than the proposed method.
 In all loss functions, we used SNR instead of the original SI-SDR to retain the scale of signals for the mixing operation.

\begin{figure}
  \centering
  \centerline{\includegraphics[width=0.5\columnwidth]{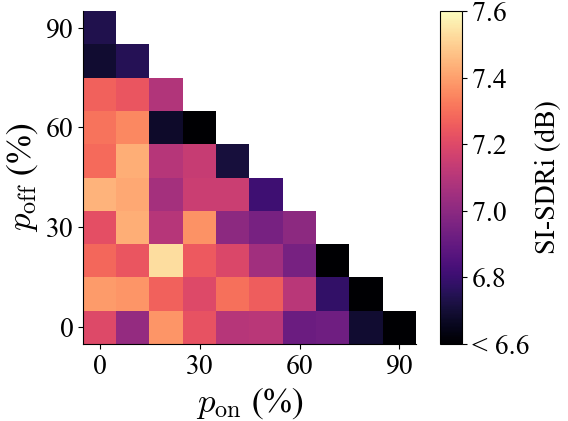}}
\vspace{-3mm}
\caption{
    Grid search of $p_{\mathrm{on}}$ and $p_{\mathrm{off}}$
    for Proposed\_A on the conditions with environmental sound noise (``noise").
}
\label{fig:MS_grid}
\vspace{-3mm}
\end{figure}

\subsection{Experimental results}
\vspace{-1mm}

Table~\ref{tab:ablation} shows the experimental results on the conditions with environmental sound noise and with an interfering speaker. 
Proposed\_A outperformed Baseline\_A in SI-SDRi and SDRi, which indicates that our direct estimation method works well for accurate estimation.
Our method can improve performance only by changing the visual encoder to a state-of-the-art encoder.
Then, Proposed\_r outperformed even the combination of the state-of-the-art methods, Baseline\_r.
Further, the ablation studies on the attention mechanism and muting strategy 
 show that each method can improve the performance of our framework although the improvement is small in Proposed\_r.
  Figure~\ref{fig:MS_grid} shows that the muting strategy tends to be effective only when $p_{\mathrm{on}} + p_{\mathrm{off}}$ is not so large.
 
The computational efficiency of the proposed method is shown in the rightmost column of Table~\ref{tab:ablation}.
As we adopted the single-path framework (single sequence of the audio encoder, TCN stacks, and audio decoder) unlike the dual-path baseline framework, the number of parameters of Proposed\_A is fewer than that of Baseline\_A by 16\%.
Proposed\_r is significantly lightweight compared to Baseline\_r because the reentry and SpEx++ have modules that do not appear in our simple network for their state-of-the-art performance.
The rightmost column also shows that
we can use the attention mechanism with a very slight parameter increase
 and apply the muting strategy without increasing a single model parameter. 
Table~\ref{tab:3_speakers} compares the baseline and the proposed method 
 on the more complex interfering speech conditions.
Here again, our method outperformed the combination of the state-of-the-art methods.

{\tabcolsep=3mm
\begin{table}[t]
\centering
\caption{
Evaluation under complex interfering conditions.
}
\label{tab:3_speakers}
\vspace{-2mm}
\begin{tabular}{cccc}
\toprule
Interference & Methods  & SI-SDRi (dB) $\uparrow$  & SDRi (dB) $\uparrow$\\
\midrule
\multirow{4}{*}{noise+spk} & Baseline\_A
& 6.17 & 6.70        \\
& Proposed\_A & 6.35 & 6.96 \\
& 
 Baseline\_r & 7.34 &  7.91    \\
& Proposed\_r & \textbf{7.53} & \textbf{8.14}\\
\midrule
\multirow{4}{*}{2spk} & 
Baseline\_A & 6.28 & 6.73        
\\
& Proposed\_A & 5.66 & 6.19\\
& 
Baseline\_r & 7.46 &  7.94   \\
& Proposed\_r & \textbf{7.59} & \textbf{8.08}\\
\bottomrule
\end{tabular}
\vspace{-4mm}
\end{table}
}

\vspace{-0.5mm}
\section{Conclusion}
\label{sec:conclusion}

This paper presented an AV-SE method that estimates the mixture of on-screen and off-screen target speech from noisy audio.
We fused two multimodal clues to extract the target mixture in a computationally efficient manner.
We also introduced the attention mechanism and proposed the muting strategy
 to improve the performance of our model further.
We experimentally confirmed that our method estimated the target mixture more accurately and efficiently
 compared to the baseline method.
Our future work includes the evaluation of our method using more realistic data.
We also plan to extend the proposed method to extract non-speech off-screen target sounds, such as alarms and sirens.

\vspace{12pt}

\end{document}